\documentclass[prd, preprint, 11pt]{revtex4-1}

\usepackage{amsmath}
\usepackage{amssymb}
\usepackage{setspace}
\usepackage{graphicx}
\usepackage{natbib}
\usepackage{float}

\begin{document}
 
 %

\begin{center}
 { \large {\bf Cosmological Constant, Quantum Measurement, and the Problem of Time}}


\vskip 0.3 in

{\large{\bf Shreya Banerjee, Sayantani Bera and Tejinder P.  Singh}}

{\it Tata Institute of Fundamental Research,}
{\it Homi Bhabha Road, Mumbai 400005, India}\\
\bigskip
{\tt shreya.banerjee@tifr.res.in, sayantani.bera@tifr.res.in, tpsingh@tifr.res.in}\\

\end{center}

\bigskip
\bigskip

\centerline{\bf ABSTRACT}

\noindent Three of the big puzzles of theoretical physics are the following: (i) There is apparently no time evolution in the dynamics of quantum general relativity, because the allowed quantum states must obey the Hamiltonian constraint. (ii) During a quantum measurement, the state of the quantum system randomly collapses from being in a linear superposition of the eigenstates of the measured observable, to just one of the eigenstates, in apparent violation of the predictions of the deterministic, linear Schr\"{o}dinger equation. (iii) The observed value of the cosmological constant is exceedingly small, compared to its natural value, creating a serious fine-tuning problem.  In this essay we propose a novel idea to show how the three problems help solve each other.

\noindent 

\vskip 1 in

\centerline{March 26, 2015}

\bigskip

\centerline{This essay received an honourable mention in the Gravity Research Foundation 2015 Essay Contest}

\bigskip

\bigskip


\newpage

\setstretch{1.1}

\centerline{\large\it Three Big Problems of Theoretical Physics} 

\bigskip

\noindent {\bf 1. The Cosmological Constant Puzzle:} In the standard $\Lambda CDM$ model of cosmology, the cosmological constant $\Lambda$ takes a non-zero value, comparable to the present matter density. This value is some $120$ orders of magnitude smaller than its natural value arising from a Planck scale cut-off of the vacuum energy density, creating a serious fine-tuning problem. Why should $\Lambda$ be so small, and yet non-zero? And what is so special about the present  epoch, that $\Lambda$ and the matter density are of the same order now (the cosmic coincidence problem)?


\bigskip

\noindent {\bf 2. The Quantum Measurement Problem:} In quantum mechanics, during a measurement, the wave-function of the quantum system apparently collapses from being in a superposition of the eigenstates of the measured observable, to being in only one of the eigenstates. Why does this breakdown of superposition take place, even though the Schr\"{o}dinger equation, which is deterministic and linear, does not predict the collapse? The outcome of a measurement is random, with its probability being given by the Born rule. Why is the outcome probabilistic, even though the initial condition is known exactly, and there is no sampling space to choose from? 


\bigskip

\noindent{\bf 3.  The Problem of Time in Quantum Gravity:} Classical general relativity is a reparametrization invariant theory, which results in the theory obeying the Hamiltonian constraint and the diffeomorphism constraint. In the Dirac quantisation of the theory, the allowed wave-functionals must obey these constraints, and in particular the Hamiltonian constraint converts the functional Schr\"{o}dinger  equation of the theory in to the timeless Wheeler-DeWitt equation:
\begin{equation}
i\hbar \frac{\delta\Psi}{\delta\tau} = \hat{H}\Psi = 0
\end{equation}
The Hamiltonian constraint creates a problem for the quantum theory: how does one understand evolution of the dynamics, and how does one recover the time evolution of the classical theory, starting from a timeless dynamics?


\bigskip
\bigskip

\centerline{\large\it How these Three Problems Solve Each Other} 

\medskip

\noindent In this essay we propose that these three problems are inter-connected, and help solve each other. Let us begin with the quantum measurement problem. A phenomenological modification of the non-relativistic Schr\"{o}dinger equation, known as Continuous Spontaneous Localisation  [CSL] has been proposed in the literature \cite{Pearle:89,Ghirardi2:90}. CSL modifies quantum mechanics to include an anti-hermitian, norm-preserving,  stochastic and non-linear component in the Hamiltonian in the Schr\"{o}dinger equation. For a system of particles, the CSL equation is given by
\begin{eqnarray} \label{eq:csl-massa}
d\psi_t  =   \left[-\frac{i}{\hbar}\hat{H}dt 
 + \frac{\sqrt{\gamma}}{m_{0}}\int d\mathbf{x} (M(\mathbf{x}) - \langle M(\mathbf{x}) \rangle_t)
dW_{t}(\mathbf{x}) \right. \nonumber \\
 -  \left. \frac{\gamma}{2m_{0}^{2}} \int d\mathbf{x}\, d\mathbf{y}\, g({\bf x} - {\bf y})
(M(\mathbf{x}) - \langle M(\mathbf{x}) \rangle_t) (M(\mathbf{y}) - \langle M(\mathbf{y}) \rangle_t)dt\right] \psi_t  
\end{eqnarray}
Here $\hat{H}$ is the standard quantum Hamiltonian operator of the system, and the two new anti-hermitian terms induce collapse of the wave-function in space. $m_0$ is a reference mass, usually taken as nucleon mass, and the parameter $\gamma$ sets the strength of the collapse process. $M({\bf x})$ is the mass density operator, $g({\bf x} - {\bf y})$ is a smearing function, and $W_t({\bf x})$ is an ensemble of independent Wiener processes.  The wave-function is now a stochastic entity, and one can construct a Lindblad type linear master equation for the corresponding density matrix.


The strength parameter $\gamma$ is linearly proportional to the total mass of the system. As a consequence of CSL, quantum linear superposition becomes an approximate principle of nature. For microscopic systems, superposition lifetimes are astronomically large, and the theory reduces to the Schr\"{o}dinger equation. For macroscopic systems superposition lifetimes are extremely small, which explains the collapse of the wave-function during a quantum measurement, in accordance with the Born rule, and for such systems the theory reduces to classical mechanics. CSL is falsifiable and for mesoscopic systems the departures from quantum theory are testable in the laboratory, and the theory is indeed being subjected to rigorous experimental tests in ongoing experiments \cite{RMP:2012}.


What are the likely implications of CSL for quantum general relativity [QGR]? We can envisage that in QGR too, there will appear a CSL type anti-Hermitian stochastic modification $iH_{st}$ to the functional 
Schr\"{o}dinger equation, converting the Wheeler-DeWitt equation in to a stochastic equation: 
\begin{equation}
i\hbar \frac{\delta\Psi}{\delta\tau} = \hat{H}\Psi + i H_{st}\Psi
\label{wds}
\end{equation}
Allowed quantum states must still obey the Dirac constraint $\hat{H}\Psi =0$, and hence, in addition to this stochastic Hamiltonian constraint we now have the stochastic evolution equation
\begin{equation}
i\hbar \frac{\delta\Psi}{\delta\tau} =  i H_{st}\Psi
\label{qgs}
\end{equation}
Evolution has been restored in the dynamics! And the inspiration has come from resolution of the quantum measurement problem. More importantly perhaps, the anti-Hermitian stochastic part causes the collapse (more appropriately localisation) in superspace, to a dynamics which then evolves classically, following the laws of classical general relativity.


For ease of visualisation, and simplicity of presentation, let us see this worked out in a minisuperspace example. We will then also see how a non-zero cosmological constant naturally enters the picture, as an inevitable candidate for the CSL strength parameter $\gamma$, defined above. Our minisuperspace model will consist of a homogeneous massive scalar field $\phi(t)$ in a spatially flat Friedmann-Roberston-Walker universe described by a scale factor $a(t)$. The Wheeler-DeWitt equation is given by
\cite{Kiefer:2012}
\begin{equation}
\hat{H}\Psi \equiv \frac{1}{2} \left( \frac{\hbar^2}{a^2}\frac{\partial\ }{\partial a} \left(a\frac{\partial\ }{\partial a}\right) - \frac{\hbar^2}{a^3}  \frac{\partial^2\ }{\partial \phi^2} -a   +
\frac{\Lambda a^3}{3}  + m^2 a^3 \phi^2   \right) \Psi (a, \phi) = 0 
\label{ap}
\end{equation}

Following Eqns. (\ref{wds}) and (\ref{qgs})  the CSL type stochastic differential equation for this model takes the form
\begin{equation}
d\Psi (a, \phi) = \left[ \sqrt{\gamma_1} ( a - \langle a \rangle ) dW_1 - 
\frac{\gamma_1}{2}(a-\langle a \rangle)^{2} dt  + \sqrt{\gamma_2} ( \phi - \langle \phi \rangle ) dW_2 - 
\frac{\gamma_2}{2}(\phi-\langle \phi \rangle)^{2} dt \right] \Psi (a, \phi)
\label{aps}
\end{equation} 
where $\gamma_1$ and $\gamma_2$ are the collapse strength parameters for the gravity part (the scale factor) and the matter part (the scalar field) respectively. Solutions of Eqn. (\ref{ap}) evolve according to Eqn. (\ref{aps}). Interestingly, the picture now resembles quantised gauge theories of particle physics: there is quantum evolution, and there are gauge constraints; and the evolution equation is no longer mixed up with the constraints.


In order to arrive at a realistic description of the universe, we assume that the gravitational rate 
constant $\gamma_1$ far exceeds the matter rate constant $\gamma_2$: $\gamma_1 \gg \gamma_2$, so that to leading approximation we may write this equation as
\begin{equation}
d\Psi (a, \phi) = \left[ \sqrt{\gamma_1} ( a - \langle a \rangle ) dW_1 - 
\frac{\gamma_1}{2}(a-\langle a \rangle)^{2} dt    \right] \Psi (a, \phi)
\label{appst}
\end{equation} 
The relative values of the two constants ensure that gravity becomes classical before matter does, since the collapse time for the scale factor ($\gamma_1^{-1}$) is much less than that for the scalar field 
($\gamma_2^{-1}$). The collapse of the wave function causes one particular classical geometry to be picked up, around which the wave function is localised, and the collapse obeys the Born probability rule.
The approximation (\ref{appst}) is valid for times less than $\gamma_{2}^{-1}$, and for $t\sim \gamma_2^{-1}$ the terms involving collapse of the scalar field also come into play.


Now $\gamma_1$ can be taken as a new constant of nature, but if we introspect we realise it is possible to relate $\gamma_1$ to the one free parameter available in gravitational physics, namely the cosmological constant $\Lambda$. In the original CSL model, the collapse parameter $\gamma$ scales with the mass of the system as
\begin{equation}
\gamma = \frac{m}{m_0}\gamma_0
\end{equation}
with $\gamma_0$ being the value of the collapse rate constant for a nucleon having mass $m_0$. This ascertains that more massive systems collapse much faster, compared to microscopic ones. In analogy to this, we propose that the gravitational collapse strength parameter is related to the value $\Lambda$ of the cosmological constant in our universe:
\begin{equation}
\gamma_1 = \frac{E_\Lambda}{E_{Pl}}\gamma_{Pl}
\label{gamma1} 
\end{equation} 
The assumed linearity of $\gamma_1$ with energy is motivated from the original CSL model, which has proportionality with mass, and the energy scale $E_\Lambda$ is determined as usual from the energy density: $E_\Lambda = \rho_\Lambda^{1/4}$. It is assumed that at Planck energy $E_{Pl}=10^{19}$ GeV the rate constant has the Planck value $\gamma_{Pl}=10^{43}$ sec$^{-1}$, implying that if $\Lambda$ was at the Planck scale classicalization of gravity would happen at Planck time.


The assumed dependence of $\gamma_1$ leads to some important consequences. It explains why the cosmological constant must be non-zero. For if it were to be zero, the time $\gamma_1^{-1}$ taken for gravity to become classical would be infinite, which is of course unacceptable. 

Next, we know that gravity should have become classical in the early universe before the epoch corresponding to the electroweak symmetry breaking energy scale, say before about 1 TeV ($10^{-12}$ sec).  Also, assuming gravity to be classical during inflation, the latest epoch that inflation could have possibly occurred and yet solved horizon and flatness problems, is around 1 TeV (for instance, fast conformal inflation with about 37 e-folds \cite{Dimopoulos:2006,Ringeval:2010}). So if we take $\gamma_1^{-1}$ to be $10^{-12}$ sec in Eqn. (\ref{gamma1}) we get 
$\rho_{\Lambda}\approx 10^{-47}$ GeV$^4$, which is the observed value of the cosmological constant!
We conclude that the astronomically inferred value of $\Lambda$ is the minimum value needed for a sufficiently early classicalization of gravity in the early universe. A smaller value of $\Lambda$ would contradict established physics of the early universe.


A larger value of $\Lambda$ is in principle permitted, but would contradict cosmological data. Thus we predict from our work that the transition from quantum to classical gravity occurs around 1 TeV scale, and is followed by an inflationary epoch around the same energy scale. The transition from quantum to classical matter occurs during the epoch of inflation via the CSL mechanism, as has been recently proposed by various authors, and this leaves an imprint on the cosmic microwave background
\cite{Das:2013,Das:2014,Martin:2012,Unanue:08,Leon:2015}.


The problem of time in quantum gravity, the cosmological constant puzzle, and the quantum measurement problem, a priori do not seem to have anything to do with each other. However, we have seen that the same CSL mechanism which solves the measurement problem, also explains how a classical universe with time evolution emerges from timeless quantum general relativity. It is natural to assume that the rate constant for stochastic classicalization of gravity is proportional to the energy scale set by the cosmological constant, because there is no other free parameter in the theory. Doing so then remarkably explains why the cosmological constant is not zero, and we show that it takes the minimal non-zero value consistent with known physics of the early universe.

\bigskip

This work is supported by a grant from the John Templeton Foundation (\#39530).


\bigskip

\bigskip

\centerline{\bf REFERENCES}

\bibliography{biblioqmts3}

\end{document}